\documentclass[ps2pdf]{myelsart}

\usepackage{amssymb}     
\usepackage[usenames]
           {color}       
\usepackage{comment}     
\usepackage{enumerate}   
\usepackage{epsfig}      
\usepackage{graphicx}    
\usepackage{latexsym}    
\usepackage{multicol}    
\usepackage{pdflscape}   
\usepackage{subfigure}   
\usepackage{url}         

\usepackage[
  linktocpage=true,      
  colorlinks=true,       
  pdftitle={Approximation Algorithms for Shortest Descending Paths in
    Terrains},
  pdfauthor={Ahmed et al.},
  pdfsubject={Computer Science}
  ]{hyperref}

\newtheorem{obs}[thm]{Observation}

\begin{document}

\begin{frontmatter}

  \title{Approximation Algorithms for Shortest Descending
    Paths in Terrains\thanksref{NSERC}}
  \thanks[NSERC]{Research partially supported by NSERC}
  \author[UW]{Mustaq Ahmed}
  \ead{m6ahmed@uwaterloo.ca}
  \author[ISI]{Sandip Das}
  \ead{sandipdas@isical.ac.in}
  \author[TCS]{Sachin Lodha}
  \ead{sachin.lodha@tcs.com}
  \author[UW]{Anna Lubiw}
  \ead{alubiw@uwaterloo.ca}
  \author[CU]{Anil Maheshwari}
  \ead{anil@scs.carleton.ca}
  \author[TCS]{Sasanka Roy}
  \ead{sasanka.roy@tcs.com}
  
  \address[UW]{
    David R. Cheriton School of Computer Science,
	 University of Waterloo,
    Waterloo, ON, N2L 3G1, Canada
  }
  \address[ISI]
  {
    Indian Statistical Institute,
    Kolkata, India
  }
  \address[TCS]
  {
    Tata Consultancy Services Ltd.,
    Pune, India
  }
  \address[CU]{
    School of Computer Science,
    Carleton University,
    Ottawa, ON, K1S 5B6, Canada
  }

  \begin{abstract}
    %
%
%

%

A path from $s$ to $t$ on a polyhedral terrain is \emph{descending} if
the height of a point $p$ never increases while we move $p$ along the
path from $s$ to $t$. No efficient algorithm is known to find a
shortest descending path (SDP) from $s$ to $t$ in a polyhedral
terrain. We give two approximation algorithms (more precisely, FPTASs)
that solve the SDP problem on general terrains. Both algorithms are
simple, robust and easy to implement.

  \end{abstract}


  
\end{frontmatter}

%
%
%
%

%

\section{Introduction}
\label{L2:Intro}

Finding a shortest path between two points in a geometric domain is
one of the most fundamental problems in computational geometry. One
extensively-studied version of the problem is to compute a shortest
path on a polyhedral terrain; this has many applications in robotics,
industrial automation, Geographic Information Systems and wire
routing. Our paper is about a variant of this problem for which no
efficient algorithm is known, the \emph{Shortest Descending Path (SDP)
Problem\/}: given a polyhedral terrain, and points $s$ and $t$ on the
surface, find a shortest path on the surface from $s$ to $t$ such
that, as a point travels along the path, its elevation, or
$z$-coordinate, never increases. We need to compute a shortest
descending path, for example, for laying a canal of minimum length
from the source of water at the top of a mountain to fields for
irrigation purpose, and for skiing down a mountain along a shortest
route~\cite{Ahmed.06,Roy.07}.

The SDP problem was introduced by de Berg and van
Kreveld~\cite{Berg.97}, who gave a polynomial time algorithm to decide
existence of a descending path between two points. Since then the
problem has been studied in different restricted
settings~\cite{Ahmed.06,Ahmed.07b,Roy.07} (See
Section~\ref{L3:Back.Related} for a brief survey), but the SDP problem
on \emph{general\/} terrains remained open in the sense that neither a
polynomial time algorithm nor a polynomial time approximation scheme
(PTAS) was known. In this paper we present two approximation
algorithms (more precisely, fully polynomial time approximation
schemes, FPTASs) to find SDPs in general terrains. These algorithms
have appeared in preliminary forms in Ahmed and Lubiw~\cite{Ahmed.07}
and in Roy et al.~\cite{Roy.07b} respectively. Both the algorithms
discretize the terrain by adding Steiner points along the edges, thus
transforming the geometric shortest path problem into a combinatorial
shortest path problem in a graph. This approach has been used before
for related shortest path problems such as the Weighted Region Problem
and the Shortest Anisotropic Path Problem (discussed in
Section~\ref{L3:Back.Related}). In those results, Steiner points are
placed \emph{independently\/} along each edge. Such independent
placement fails for SDPs, and our main new ingredient is to place
Steiner points by slicing the terrain with horizontal planes. Both the
algorithms presented here are simple, robust and easy to
implement.

In our first algorithm, given a vertex $s$ in a triangulated terrain,
and a constant $\epsilon \in (0,1]$, we discretize the terrain with
\(
  O \left( \frac{n^2 X}{\epsilon} \right)
\)
Steiner points so that after an
  \(
    O \left(
      \frac{n^2 X}{\epsilon} \log \left( \frac{n X}{\epsilon} \right)
    \right)
  \)
-time preprocessing phase, we can determine a
$(1+\epsilon)$-approximate SDP from $s$ to any point $v$ in $O(n)$
time if $v$ is either a vertex of the terrain or a Steiner point, and
in
\(
  O \left( \frac{n X}{\epsilon} \right)
\)
time otherwise, where $n$ is the number of vertices of the terrain,
and $X$ is a parameter of the geometry of the terrain. More precisely,
$X = \frac{L}{h} \cdot \frac{1}{\cos\theta} = \frac{L}{h} \sec\theta$,
where $L$ is the length of the longest edge, $h$ is the smallest
distance of a vertex from a non-adjacent edge in the same face
(i.e.~the smallest 2D height of a triangular face), and $\theta$ is
the largest acute angle between a non-level edge and a vertical
line. Our second algorithm places Steiner points in a different
manner, which modifies the above preprocessing time and the two query
times to
  \(
    O \left(
      \frac{n^2 X'}{\epsilon} \log^2 \left( \frac{n X'}{\epsilon} \right)
    \right)
  \),
$O(n)$, and
\(
  O \left( \frac{n X'}{\epsilon}
    \log \left( \frac{n X'}{\epsilon} \right)
  \right)
\)
respectively, where $X' = \frac{L}{h}$. In comparison, the first
algorithm is faster in terms of $n$, $\epsilon$ and $\frac{L}{h}$, but
it depends heavily on the inclination of the non-level edges. On the
other hand, the second algorithm does not depend at all on edge
inclinations, and hence is better for terrains with almost level
edges. It is straightforward to follow a ``hybrid'' approach that
first checks the edge inclinations of the input terrain, and then runs
whichever of these two algorithms ensures a better running time for
that particular terrain.

The paper is organized as follows. In Section~\ref{L2:Background} we
define a few terms, discuss the properties of SDPs, and mention
related results. Sections~\ref{L2:Steiner} and~\ref{L2:Steiner3} give
details of our approximation algorithms. We conclude in
Section~\ref{L2:Conclusion} with a few open problems.

\section{Preliminaries}
\label{L2:Background}

\subsection{Terminology}
\label{L3:Back.Term}

A terrain is a 2D surface in 3D space with the property that every
vertical line intersects it in at most one point~\cite{Berg.00}. We
consider triangulated terrains. For any point $p$ in the terrain,
$h(p)$ denotes the height of $p$, i.e.,~the $z$-coordinate of $p$. We
assume without loss of generality that all points of the terrain lie
above the plane $z = 0$. An edge or face in 3D is \emph{level\/} if
all points on that edge or face have the same height. We add $s$ as a
vertex of the terrain. Let $n$ be the number of vertices in the
terrain. By Euler's formula~\cite{Berg.00}, the terrain has at most $3
n$ edges, and at most $2 n$ faces.

\begin{figure}[htb]
  \hspace*{\fill}
  \input{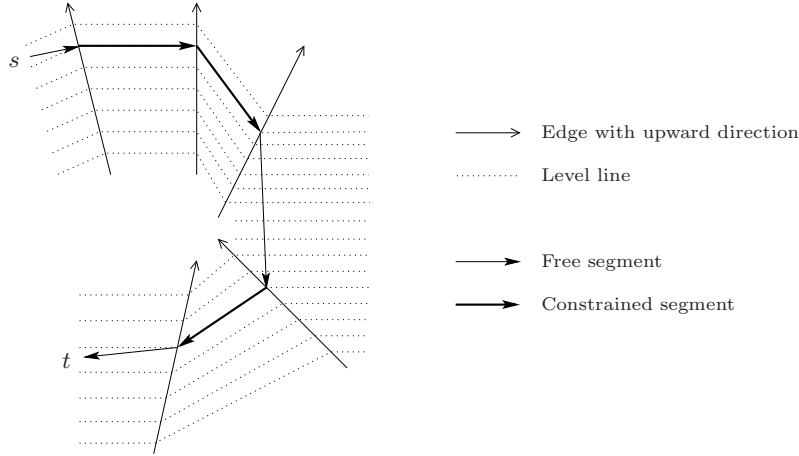}
  \hspace*{\fill}
  \caption{General legend for the figures in this
    paper\label{fig:Legend}}
\end{figure}

We reserve the terms ``edge'' and ``vertex'' for features of the
terrain. We use the term ``segment'' to denote a line segment of a
path, and ``node'' to denote an endpoint of a segment. We use ``node''
and ``link'' to mean the corresponding entities in a graph or a
tree. Figure~\ref{fig:Legend} shows the convention we will use in our
figures to mark various components related to a descending path. In
particular, an arrow with a solid, dark arrowhead denotes a path
segment, and the arrow may be heavy to mark a level segment. In the
figures where the direction of the edges are important, we again use
arrows to mark the upward direction, but we make the arrowheads
V-shaped (``open'') in this case to differentiate the edges from the
segments. Dotted lines are used to show level lines in a face.

A path $P$ from $s$ to $t$ on the terrain is \emph{descending\/} if
the $z$-coordinate of a point $p$ never increases while we move $p$
along the path from $s$ to $t$. We assume that all paths and segments
in our discussion are directed. Our discussion relies on the following
known~\cite{Ahmed.06,Roy.07} properties of an SDP:

\begin{lem} \label{lem:LSDP.OptSubpath}
  Any subpath of an SDP is an SDP.
\end{lem}

\begin{lem} \label{lem:LSDP.Deflection}
  An unfolded SDP is not always a straight line segment.
\end{lem}

\begin{lem} \label{lem:SDP.OneLinePerFace}
  The intersection of an SDP $P$ with a face of the terrain is either
  empty or a line segment.
\end{lem}

\subsection{Related Work}
\label{L3:Back.Related}

The SDP problem was introduced by de Berg and van
Kreveld~\cite{Berg.97}, who gave an algorithm to preprocess a terrain
in $O(n\log n)$ time so that it can be decided in $O(\log n)$ time if
there exists a descending path between any pair of vertices. They did
not consider the length of the path, and left open the problem of
finding the shortest such path. Roy, Das and Nandy~\cite{Roy.07}
solved the SDP problem for two special classes of terrains. For convex
(or concave) terrains, they use the continuous Dijkstra approach to
preprocess the terrain in $O(n^2 \log n)$ time and $O(n^2)$ space so
that an SDP of size $k$ can be determined in $O(k + \log n)$ time. For
a terrain consisting of edges parallel to one another, they find an
SDP in $O(n \log n)$ time by transforming selected faces of the
terrain in a way that makes the unfolded SDP a straight line segment.
Roy~\cite{Roy.08} has recently improved this running time to $O(n)$,
by replacing a sorting step in the previous algorithm with a
divide-and-conquer technique. Ahmed and Lubiw~\cite{Ahmed.06} examined
the basic properties of SDPs that show the similarities and the
dissimilarities between SDPs and shortest paths, and indicated why a
shortest path algorithm like the continuous Dijkstra approach cannot
be used directly to solve the SDP problem on general terrains. They
also gave an $O(n^{3.5} \log(\frac{1}{\epsilon}))$ time algorithm that
finds a $(1+\epsilon)$-approximate SDP through a \emph{given\/}
sequence of faces. Their algorithm first formulates the problem as a
convex optimization problem, which is then solved using a standard
technique in convex programming. In a more recent work Ahmed and
Lubiw~\cite{Ahmed.07b} gave a full characterization of the bend angles
of an SDP, which shows that the bend angles along an SDP follows a
generalized form of Snell's law of refraction of light. This result
implies that computing an exact SDP is not easy even when we know the
sequence of faces used by the SDP, due to numerical issues similar to
the ones faced by Mitchell and Papadimitriou while computing a
shortest path in the Weighted Region
Problem~\cite[Section~8]{Mitchell.91}.

It was Papadimitriou~\cite{Papadimitriou.85} who first introduced the
idea of discretizing space by adding Steiner points and approximating
a shortest path through the space by a shortest path in the graph of
Steiner points. He did this to find a shortest obstacle-avoiding path
in 3D---a problem for which computing an exact solution is
NP-hard~\cite{Canny.87}. On polyhedral surfaces, the Steiner point
approach has been used in approximation algorithms for many variants
of the shortest path problem, particularly those in which the shortest
path does not unfold to a straight line segment. One such variant is
the Weighted Region Problem~\cite{Mitchell.91}. In this problem, a set
of constant weights is used to model the difference in costs of travel
in different regions on the surface, and the goal is to minimize the
weighted length of a path. Mitchell and
Papadimitriou~\cite{Mitchell.91} used the continuous Dijkstra approach
to get an approximate solution in
\(
  O\left(
    n^8 \log\left(\frac{n}{\epsilon}\right)
  \right)
\)
time. Following their result, several faster approximation
schemes~\cite{Aleksandrov.98,Aleksandrov.00,Aleksandrov.05,Cheng.07,Sun.06a}
have been devised, all using the Steiner point approach. The Steiner
points are placed along the edges of the terrain, except that
Aleksandrov et al.~\cite{Aleksandrov.05} place them along the
bisectors of the face angles. A comparison between these algorithms
can be found in Aleksandrov et al.~\cite{Aleksandrov.05}.

One generalization of the Weighted Region Problem is finding a
shortest anisotropic path~\cite{Rowe.90}, where the weight assigned to
a region depends on the direction of travel. The weights in this
problem capture, for example, the effect the gravity and friction on a
vehicle moving on a slope. Lanthier et al.~\cite{Lanthier.99}, Sun and
Reif~\cite{Sun.05} and Sun and Bu~\cite{Sun.06b} solved this problem
by placing Steiner points along the edges.

Note that all the above-mentioned Steiner point approaches place the
Steiner points in a face without considering the Steiner points in the
neighboring faces. This strategy works because we can travel between
any two points in a face. In the case of shortest anisotropic paths,
the straight-line path may be in a forbidden direction, but it is
almost always assumed that the allowed directions permit a zigzag path
to any destination (like tacking against the wind in a sailboat). The
one exception is that Sun and Reif~\cite{Sun.05} consider the
Anisotropic Path Problem where a set of non-adjacent faces have
directions that are unreachable even with zigzagging. Their solution
involves propagating extra Steiner points across each of these
partially traversable faces. For the SDP problem, ascending directions
are unreachable in \emph{every\/} face, which necessitates our
non-local strategy of placing Steiner points.

\subsection{The Bushwhack Algorithm}
\label{L3:Back.Bushwhack}

To compute a shortest path in the graph of Steiner points in a terrain
we use a variant of Dijkstra's algorithm developed by Sun and
Reif~\cite{Sun.01}. Their algorithm, called the Bushwhack algorithm,
achieves $O(|V| \log |V|)$ running time by utilizing certain geometric
properties of the paths in such a graph. The algorithm has been used
in shortest path algorithms for the Weighted Region
Problem~\cite{Aleksandrov.05,Sun.06a} and the Shortest Anisotropic
Path Problem~\cite{Sun.05}.

\begin{figure}[htb]
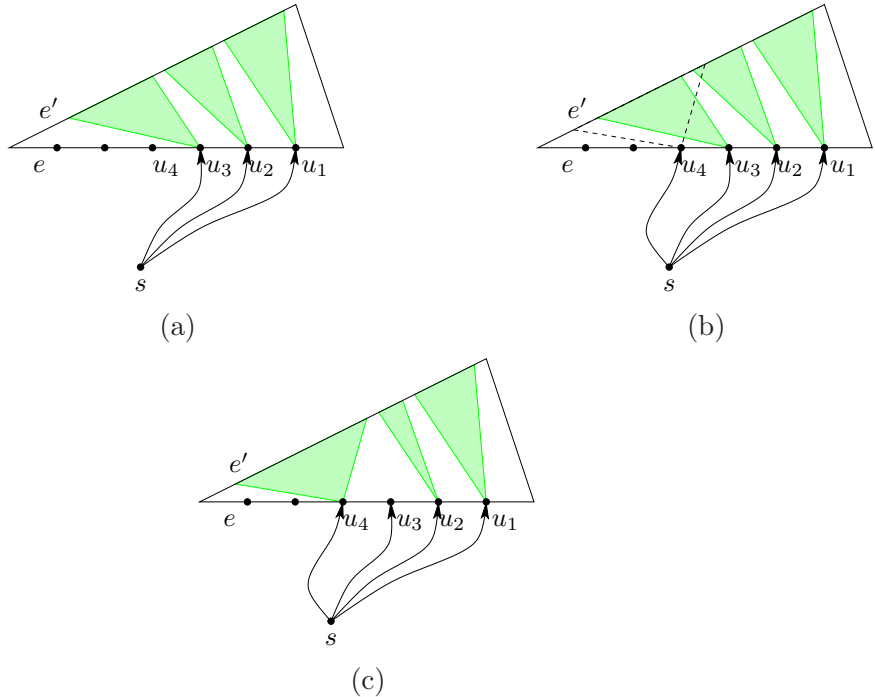

  \hspace*{\fill}
  \subfigure[]{
    \input{fig/Others-Sun06-1.pstex_t}
    \label{fig:Others-Sun06-1}
  }
  \hspace*{\fill}
  \subfigure[]{
    \input{fig/Others-Sun06-2.pstex_t}
    \label{fig:Others-Sun06-2}
  }
  \hspace*{\fill}
  \subfigure[]{
    \input{fig/Others-Sun06-3.pstex_t}
    \label{fig:Others-Sun06-3}
  }
  \hspace*{\fill}
  \caption{Maintaining the list $I_{e,e'}$ in the Bushwhack algorithm}
  \label{fig:Others-Sun06}
\end{figure}

The Bushwhack algorithm relies on a simple, yet important, property of
shortest paths on terrains: two shortest paths through different face
sequences do not intersect each other at an interior point of a
face. As a result, for any two consecutive Steiner points $u_1$ and
$u_2$ on edge $e$ for which the distances from $s$ are already known,
the corresponding sets of ``possible next nodes on the path'' are
disjoint, as shown using shading in Figure~\ref{fig:Others-Sun06-1}.
This property makes it possible to consider only a subset of links at
a Steiner point $v$ when expanding the shortest path tree onwards from
$v$ using Dijkstra's algorithm. More precisely, Sun and Reif maintain
a dynamic list of intervals $I_{e,e'}$ for every pair of edges $e$ and
$e'$ of a common face. Each point in an interval is reachable from $s$
using a shortest path through a common sequence of intermediate
points. For every Steiner point $v$ in $e$ with known distance from
$s$, $I_{e,e'}$ contains an interval of Steiner points on $e'$ that
are likely to become the next node in the path from $s$ through
$v$. The intervals in $I_{e,e'}$ are ordered in accordance with the
ordering of the Steiner points $v$ on $e$, which enables easy
insertion of the interval for a Steiner point on $e$ whose distance
from $s$ is yet unknown. For example, right after the distance of
$u_4$ from $s$ becomes known (i.e.,~right after $u_4$ gets dequeued in
Dijkstra's algorithm) as shown in Figure~\ref{fig:Others-Sun06-2}, the
interval of the Steiner points on $e'$ that are closer to $u_4$ than
to any other Steiner points on $e$ with known distances from $s$ can
be computed in time logarithmic in the number of Steiner points on
$e'$, using binary searches (Figure~\ref{fig:Others-Sun06-3}). Let
$\cal I$ denote this interval for ease of discussion. The Bushwhack
algorithm considers only the Steiner points lying in interval $\cal I$
as the possible next nodes on the path to $u_4$, while Dijkstra's
algorithm tries all the Steiner points on $e'$. Another difference
between these two algorithms is that after $u_4$ gets dequeued,
Dijkstra's algorithm enqueues each Steiner point (or sifts it upward
in the queue if it was already there) of $\cal I$. But in the
Bushwhack algorithm, only the Steiner point $u'_4 \in \cal I$ that is
nearest from $u_4$ is enqueued (or sifted upward); other Steiner
points in $\cal I$ are considered later on if necessary, one by one
and in order of their distances from $u'_4$. Since $u'_4$ can be
located in interval $\cal I$ in constant time, each iteration of the
Bushwhack algorithm takes $O(|V|)$ time, resulting in a total running
time of $O(|V| \log |V|)$.

\subsection{Placing the Steiner Points}
\label{L3:Back.Placing}

Our approximation algorithms work by first discretizing the terrain
with many Steiner points along the edges, and then determining a
shortest path in a directed graph in which each link connects a pair
of vertices or Steiner points in a face of the terrain in the
descending (more accurately, in the non-ascending) direction.
Although the idea is similar to other Steiner point approaches
discussed in Section~\ref{L3:Back.Related}, there are two aspects of
the SDP problem that make our approach quite different from previous
Steiner point approaches.

\begin{figure}[htb]
  \hspace*{\fill}
  \input{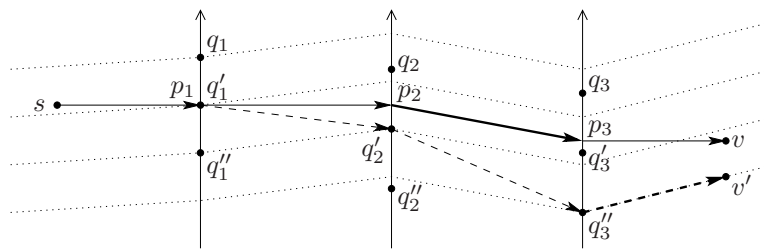}
  \hspace*{\fill}
  \caption{Problems with independently-placed Steiner points.}
  \label{fig:IndependentStPts}
\end{figure}

First, because of the nature of the SDP problem, we have to position
the Steiner points quite differently from the Steiner point approaches
discussed in Section~\ref{L3:Back.Related}. In particular, we cannot
place Steiner points in an edge without considering the heights of the
Steiner points in other edges. More elaborately, for each Steiner
point $p$ in an edge, if there is no Steiner point with height $h(p)$
in other edges of the neighboring faces, it is possible that a
descending path from $s$ to $v$ through Steiner points does not exist,
or is arbitrarily longer than the SDP. For example, consider the SDP
$P = (s, p_1, p_2, p_3, v)$ in Figure~\ref{fig:IndependentStPts},
where for each $i \in [1,3]$, $q_i$, $q'_i$ and $q''_i$ are three
consecutive Steiner points with $h(q_i) > h(q'_i) > h(q''_i)$ such
that $q_i$ is the nearest Steiner point above $p_i$. Note that in this
figure the faces have been unfolded onto a plane, and that $p_1$ and
$q'_1$ are the same point. There is no descending path from $s$ to $v$
through the Steiner points: we must cross the first edge at $q'_1$ or
lower, then cross the second edge at $q'_2$ or lower, and cross the
third edge at $q''_3$ or lower, which puts us at a height below
$h(v)$. Another important observation is that even if a descending
path exists, it may not be a good approximation of $P$. In
Figure~\ref{fig:IndependentStPts}, for example, if we want to reach
instead a point $v'$ slightly below $v$, $P'$ would be a feasible
path, but the last intermediate nodes of $P$ and $P'$ are not very
close. We can easily extend this example to an SDP $P$ going through
many edges such that the ``nearest'' descending path $P'$ gets further
away from $P$ at each step, and at one point, $P'$ starts following a
completely different sequence of edges. Clearly, we cannot ensure a
good approximation by just making the Steiner points on an edge close
to each other.

To guarantee the existence of a descending path through Steiner points
that approximates an SDP from $s$ to any vertex, we have to be able to
go through the Steiner points in a sequence of faces without ``losing
height'', i.e.,~along a level path. We achieve this by slicing the
terrain with a set of horizontal planes, and then putting Steiner
points where the planes intersect the edges. The set of horizontal
planes includes one plane through each vertex of the terrain, and
other planes in between them that are close enough to guarantee a good
approximation ratio. Our two algorithms, discussed in
Sections~\ref{L2:Steiner} and~\ref{L2:Steiner3}, differ from each
other in the manner the positions of the horizontal planes are
determined.

\begin{figure}[htb]
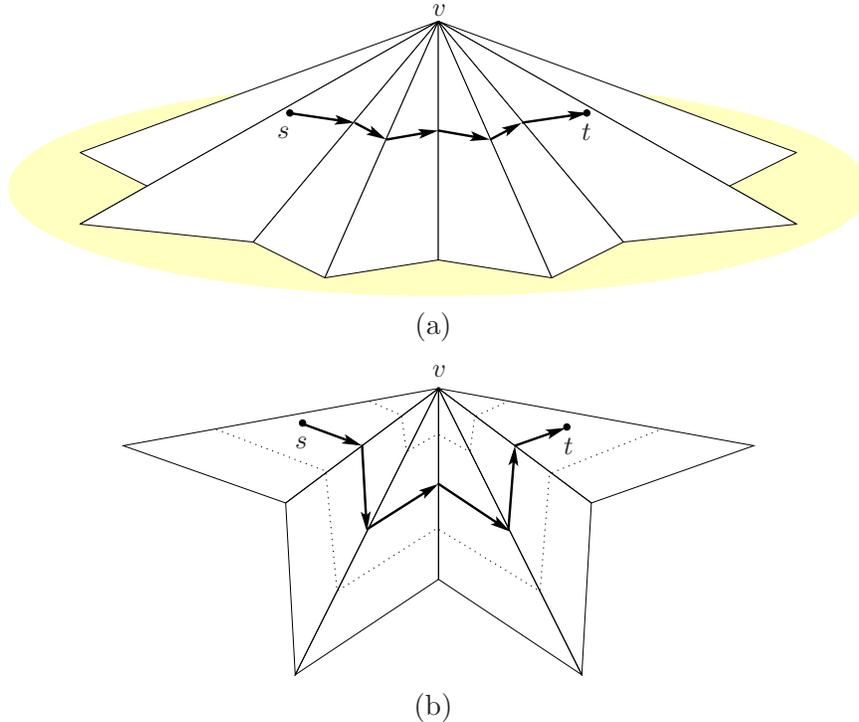

  \hspace*{\fill}
  \subfigure[]{
    \input{fig/VicinityZigzag1.pstex_t}
    \label{fig:VicinityZigzag1}
  }
  \hspace*{\fill}

  \hspace*{\fill}
  \subfigure[]{
    \input{fig/VicinityZigzag2.pstex_t}
    \label{fig:VicinityZigzag2}
  }
  \hspace*{\fill}
  \caption{An SDP that comes close to a vertex $O(n)$ number of times}
  \label{fig:VicinityZigzag}
\end{figure}

The second issue is that the previous Steiner point approaches relied
on the property that shortest paths in the Weighted Region Problem or
in the Shortest Anisotropic Path Problem cannot become very close to a
particular vertex more than once. This property does not hold for
shortest paths in the SDP problem. In fact, it is possible to
construct a terrain where an SDP becomes very close to a vertex $v$ as
many as $O(n)$ times, moving far away from $v$ after every visit of
the vicinity of $v$. Consider the terrain in
Figure~\ref{fig:VicinityZigzag1} which consists of the triangular
faces of a pyramid with a star-shaped base. The points $s$ and $t$
have the same height, so the SDP $P$ from $s$ to $t$ must consist of
level segments. Moreover, $P$ consists of $O(n)$ segments in the
figure. Figure~\ref{fig:VicinityZigzag2} shows the faces used by $P$
after unfolding them onto a plane. By moving the convex vertices at
the base away from the ``center'' of the base while keeping them on
the same plane, we can make the points of $P$ that are far away from
$v$ move even further away from $v$. Clearly it is possible to make
$P$ enter and leave a region close to $v$ as many as $O(n)$ number of
times. Because of such a possibility with an SDP, the analysis of our
Steiner point approach is completely different from previous
approaches.

%
%
%
%

%

\section{Discretizing using Uniform Steiner Points}
\label{L2:Steiner}

In our first algorithm the Steiner points on each edge are evenly
spaced. To determine their positions, we first take a set of
horizontal planes such that any two consecutive planes are within
distance $\delta$ of each other, where $\delta$ is a small constant
that depends on the approximation factor. We then put a Steiner point
at the intersection point of each of these planes with each of the
terrain edges. One important observation is that this scheme makes the
distance between consecutive Steiner points on an edge dependent on
the slope of that edge. For instance, the distance between consecutive
Steiner points is more for an almost-level edge than for an almost
vertical edge. Since $\theta$ is the largest acute angle between a
non-level edge and a vertical line, it can be shown that the distance
between consecutive Steiner points on a non-level edge is at most
$\delta \sec\theta$ (Lemma~\ref{lem:Steiner.Offset}). Because of the
situation depicted in Figure~\ref{fig:IndependentStPts}, we cannot
place extra Steiner points \emph{only\/} on the edges that are almost
level. We guarantee a good approximation ratio by choosing $\delta$
appropriately. More precisely, we make sure that $\delta \sec\theta$
is small enough for the desired approximation ratio. Note that we can
put Steiner points on a level edge without considering heights, since
a level edge can never result in the situation depicted in
Figure~\ref{fig:IndependentStPts} (because all the points in such an
edge have the same height).

\subsection{Algorithm}
\label{L3:Steiner.Alg}

Our algorithm runs in two phases. In the preprocessing phase, we place
the Steiner points, and then construct a shortest path tree in the
corresponding graph. During the query phase, the shortest path tree
gives an approximate SDP in a straightforward manner.

\subsubsection{Preprocessing Phase}
\label{L3:Steiner.Alg.Pre}

Let $\delta = \frac{\epsilon h \cos\theta}{4 n}$. We subdivide every
non-level edge $e$ of the terrain by putting Steiner points at the
points where $e$ intersects each of the following planes: $z = j
\delta$ for all positive integers $j$, and $z = h(x)$ for all vertices
$x$ of the terrain. We subdivide every level edge $e$ by putting
enough Steiner points so that the length of each part of $e$ is at
most $\delta \sec\theta$. Let $V$ be the set of all the vertices and
Steiner points in the terrain. We then construct a weighted directed
graph $G=(V,E)$ as follows, starting with $E = \emptyset$. For every
pair $(x, y)$ of points in $V$ adjacent to a face $f$ of the terrain,
we add to $E$ a directed link from $x$ to $y$ if and only if $h(x) \ge
h(y)$ and $x y$ is either an edge of the terrain or a segment through
the interior of $f$. Note that we do \emph{not\/} add a link between
two points on the same edge unless both of them are vertices. Each
link in $E$ is assigned a weight equal to the length of the
corresponding line segment in the terrain. Finally we construct a
shortest path tree $T$ rooted at $s$ in $G$ using the Bushwhack
algorithm.

Note that we are mentioning set $E$ only to make the discussion easy.
In practice, we do not construct $E$ explicitly because the neighbors
of a node $x \in V$ in the graph are determined \emph{during\/} the
execution of the Bushwhack algorithm.

\subsubsection{Query Phase}
\label{L3:Steiner.Alg.Qry}

\begin{figure}[htb]
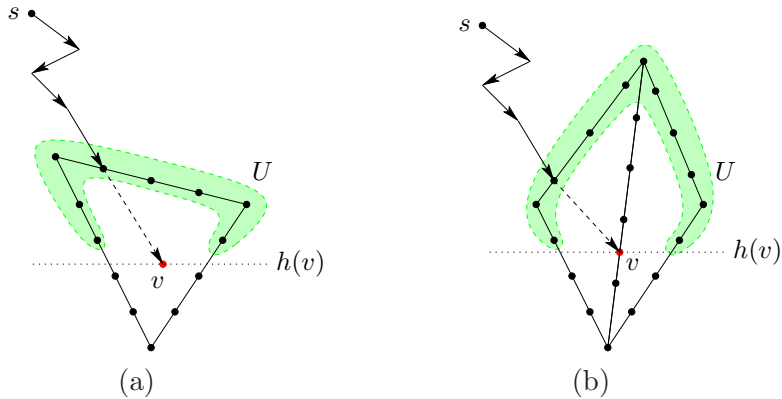

  \hspace*{\fill}
  \subfigure[]{
    \input{fig/Query-1.pstex_t}
    \label{fig:Query-1}
  }
  \hspace*{\fill}
  \subfigure[]{
    \input{fig/Query-2.pstex_t}
    \label{fig:Query-2}
  }
  \hspace*{\fill}
  \caption{Finding an SDP from $s$ to an interior point $v$ of (a) a
    face and (b) an edge}
  \label{fig:Query}
\end{figure}

When the query point $v$ is a node of $G$, we return the path from $s$
to $v$ in $T$ as an approximate SDP. Otherwise, we find the node $u$
among those in $V$ lying in the face(s) containing $v$ such that $h(u)
\ge h(v)$, and the sum of the length of the path from $s$ to $u$ in
$T$ and the length of the segment $u v$ is minimum. We return the
corresponding path from $s$ to $v$ as an approximate SDP in this
case. To elaborate more on the latter case, let $U$ be the set
consisting of the nodes $u \in V$ with the following properties:
\begin{enumerate}[(i)]
\item $u$ and $v$ lie in a common face, and
\item $h(u) \ge h(v)$.
\end{enumerate}
It is easy to see that if $v$ is an interior point of a face, then all
the nodes in $U$ lie on at most three edges of that face
(Figure~\ref{fig:Query-1}). Otherwise, $v$ is an interior point of an
edge, and there are at most four edges on which the nodes in $U$ can
lie (Figure~\ref{fig:Query-2}). Since we already know the length of an
SDP from $s$ to any $u \in U$, we can find in $|U|$ iterations the
node $u \in U$ that minimizes the length of the path constructed by
concatenating the segment $u v$ at the end of the path from $s$ to $u$
in $T$. The corresponding path is returned as an approximate SDP.

\subsection{Correctness and Analysis}
\label{L3:Steiner.Analysis}

For the proof of correctness, it is sufficient to show that an SDP $P$
from $s$ to any point $v$ in the terrain is approximated by a
descending path $P'$ such that all the segments of $P'$, except
possibly the last one, exist in $G$. We show this by constructing a
path $P'$ from $P$ in the following way. Note that $P'$ might not be
the path returned by our algorithm, but it provides an upper bound on
the length of the returned path.

Let $P = (s=p_0, p_1, p_2, \ldots, p_k, v=p_{k+1})$ be an SDP from $s$
to $v$ such that $p_i$ and $p_{i+1}$ are two different boundary points
of a common face for all $i \in [0,k-1]$, and $p_k$ and $p_{k+1}$ are
two points of a common face. For ease of discussion, let $e_i$ be an
edge of the terrain through $p_i$ for all $i \in [1,k]$ ($e_i$ can be
any edge through $p_i$ if $p_i$ is a vertex). Intuitively, we
construct $P'$ by moving each intermediate node of $P$ upward to the
nearest Steiner point. More precisely, we define a path $P' = (s=p'_0,
p'_1, p'_2, \ldots, p'_k, v=p'_{k+1})$ as follows. For each $i \in
[1,k]$, let $p'_i = p_i$ if $p_i$ is a vertex of the
terrain. Otherwise, let $p'_i$ be the nearest point from $p_i$ in $V
\cap e_i$ such that $h(p'_i) \ge h(p_i)$. Such a point always exists
in $V$ because $p_i$ is an interior point of $e_i$ in this case, and
it has two neighbors $x$ and $y$ in $V \cap e_i$ such that $h(x) \ge
h(p_i) \ge h(y)$. Note that each node of $P'$ except possibly the last
one is either a vertex or a Steiner point.

\begin{lem} \label{lem:Steiner.Feasibility}
  Path $P'$ is descending, and the part of $P'$ from $s$ to $p'_k$
  exists in $G$.
\end{lem}

\begin{pf}
  We prove that $P'$ is descending by showing that $h(p'_i) \ge
  h(p'_{i+1})$ for every $i \in [0,k]$. We have: $h(p'_i) \ge
  h(p_{i+1})$, because $h(p'_i) \ge h(p_i)$ by the definition of
  $p'_i$, and $h(p_i) \ge h(p_{i+1})$ as $P$ is descending. Now
  consider the following two cases:
  \begin{description}
  \item[Case 1:] $p'_{i+1} = p_{i+1}$ or $e_{i+1}$ is a level edge.
    In this case, $h(p'_{i+1}) = h(p_{i+1})$. It follows from the
    inequality $h(p'_i) \ge h(p_{i+1})$ that $h(p'_i) \ge
    h(p'_{i+1})$.

  \item[Case 2:] $p'_{i+1} \neq p_{i+1}$ and $e_{i+1}$ is a non-level
    edge. In this case, there is either one or no point in $e_{i+1}$
    at any particular height. Let $p''_{i+1}$ be the point in
    $e_{i+1}$ such that $h(p''_{i+1}) = h(p'_i)$, or if no such point
    exists, let $p''_{i+1}$ be the upper vertex of $e_{i+1}$. In the
    latter case, we can infer from the inequality $h(p'_i) \ge
    h(p_{i+1})$ that $h(p'_i) > h(p''_{i+1})$. Therefore we have
    $h(p'_i) \ge h(p''_{i+1})$ in both cases. Since $p''_{i+1} \in V
    \cap e_{i+1}$, the definition of $p'_{i+1}$ implies that
    $h(p''_{i+1}) \ge h(p'_{i+1})$. So, $h(p'_i) \ge h(p'_{i+1})$.

  \end{description}
  Therefore, $P'$ is a descending path.

  To show that the part of $P'$ from $s$ to $p'_k$ exists in $G$, it
  is sufficient to prove that $p'_i p'_{i+1} \in E$ for all $i \in
  [0,k-1]$, because both $p'_i$ and $p'_{i+1}$ are in $V$ by
  definition. We have already proved that $h(p'_i) \ge
  h(p'_{i+1})$. Since $p'_i$ and $p'_{i+1}$ are boundary points of a
  common face by definition, $p'_i p'_{i+1} \not\in E$ only in the
  case that both of $p'_i$ and $p'_{i+1}$ lie on a common edge, and at
  most one of them is a vertex. We show as follows that this is
  impossible. When both $p_i$ and $p_{i+1}$ are vertices of the
  terrain, both $p'_i$ and $p'_{i+1}$ are vertices. When at least one
  of $p_i$ and $p_{i+1}$ is an interior point of an edge, they cannot
  lie on a common edge~\cite[Lemma~3]{Ahmed.06}; therefore, both of
  $p'_i$ and $p'_{i+1}$ cannot lie on a common edge unless both of
  $p'_i$ and $p'_{i+1}$ are vertices. So, this is impossible that both
  $p'_i$ and $p'_{i+1}$ lie on a common edge, and at most one of them
  is a vertex. Therefore, $p'_i p'_{i+1} \in E$. \qed
\end{pf}

\begin{lem} \label{lem:Steiner.Offset}
  For all $i \in [1,k]$, $|p_i p'_i| \le \delta \sec\theta$.
\end{lem}

\begin{pf}
  \begin{figure}[htb]
    \hspace*{\fill}
    \input{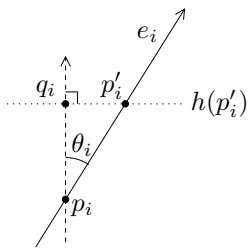}
    \hspace*{\fill}
    \caption{Bounding $|p_i p'_i|$ when $p_i \neq p'_i$ and $e_i$ is a
      non-level edge}
    \label{fig:SecTheta}
  \end{figure}

  When $p_i = p'_i$, $|p_i p'_i| = 0 < \delta \sec\theta$. When $p_i
  \neq p'_i$, and $e_i$ is a level edge, $|p_i p'_i| \le \delta
  \sec\theta$ by construction. We will now focus on the case $p_i \neq
  p'_i$ and $e_i$ is a non-level edge.

  Consider the vertical plane containing the edge $e_i$. Construct a
  line vertically upward from $p_i$ to the point $q_i$ where $h(q_i) =
  h(p'_i)$ (Figure~\ref{fig:SecTheta}). Let $\theta_i$ be the angle
  $\angle q_i p_i p'_i$. Since $h(q_i) = h(p'_i) > h(p_i)$, $\theta_i$
  is an acute angle, and hence $\theta \ge \theta_i$, which implies:
  \begin{eqnarray*}
    \cos\theta
    &\le&
    \cos\theta_i
    \,=\,
    \frac{|q_i p_i|}{|p_i p'_i|}
    \quad\Rightarrow\quad
    |p_i p'_i|
    \,\le\,
    |q_i p_i| \, \sec\theta
    \enspace
    .
  \end{eqnarray*}
  As $q_i p_i$ is a vertical line,
  \begin{eqnarray*}
    |q_i p_i|
    &=&
    h(q_i) - h(p_i)
    \,=\,
    h(p'_i) - h(p_i)
    \,\le\,
    \delta
  \end{eqnarray*}
  by construction, and therefore, $|p_i p'_i| \le |q_i p_i| \,
  \sec\theta \le \delta \sec\theta$. \qed
\end{pf}

\begin{lem} \label{lem:Steiner.ApproxFactor}
  Path $P'$ is a $(1+\epsilon)$-approximation of $P$.
\end{lem}

\begin{pf}
  When $k = 0$ implying that $P$ does not cross an edge of the
  terrain, we have $P = (s,v) = P'$ which proves the lemma trivially.
  We will now focus on the case $k > 0$.
  
  The length of $P'$ is equal to:
  \begin{eqnarray*}
    \sum_{i=0}^k |p'_i p'_{i+1}|
    &\le&
    \sum_{i=0}^k \left(
      |p'_i p_i| + |p_i p_{i+1}| + |p_{i+1} p'_{i+1}|
    \right)
    \qquad
    \mbox{(from triangle inequality)}
    \\
    &=&
    \sum_{i=0}^k |p_i p_{i+1}| + 2 \sum_{i=1}^k |p_i p'_i|
    \qquad
    \mbox{(since $p_0 = p'_0$ and $p_{k+1} = p'_{k+1}$)}
    \\
    &\le&
    \sum_{i=0}^k |p_i p_{i+1}| + 2 \sum_{i=1}^k \delta \sec\theta
    \qquad
    \mbox{(Lemma~\ref{lem:Steiner.Offset})}
    \\
    &\le&
    \sum_{i=0}^k |p_i p_{i+1}| + 2 k \delta \sec\theta
    \enspace
    .
  \end{eqnarray*}
  Because the number of faces in the terrain is at most $2 n$, and $P$
  has at most one segment in each face
  (Lemma~\ref{lem:SDP.OneLinePerFace}), we have: $k < 2 n$. Therefore,
  \begin{eqnarray*}
    \sum_{i=0}^k |p'_i p'_{i+1}|
    &<&
    \sum_{i=0}^k |p_i p_{i+1}| + 4 n \delta \sec\theta
    \\
    &=&
    \sum_{i=0}^k |p_i p_{i+1}| + \epsilon h
    \enspace
    ,
  \end{eqnarray*}
  from the definition of $\delta$. Because $k > 0$, $p_1$ lies on the
  edge opposite to $p_0$ in the face containing both $p_0$ and $p_1$,
  and therefore, $h \le |p_0 p_1| \le \sum_{i=0}^k |p_i p_{i+1}|$. So,
  \[
    \sum_{i=0}^k |p'_i p'_{i+1}|
    <
    (1+\epsilon) \sum_{i=0}^k |p_i p_{i+1}|
    \enspace
    .
  \]
  Since $P'$ is descending (Lemma~\ref{lem:Steiner.Feasibility}), it
  follows that $P'$ is a $(1+\epsilon)$-approximation of $P$. \qed
\end{pf}

\begin{lem} \label{lem:Steiner.GraphSize}
  Let $X = \left(\frac{L}{h}\right) \sec\theta$. Graph $G$ has less
  than $\frac{15 n^2 X}{\epsilon}$ nodes and $O \left( \frac{n^3
  X^2}{\epsilon} \right)$ links.  Moreover, it has less than $\frac{5
  n X}{\epsilon}$ nodes along any edge of the terrain.
\end{lem}

\begin{pf}
  We will first prove the last part of the lemma. For each edge $e$ of
  the terrain, the number of Steiner points corresponding to the
  planes $z = j \delta$ is at most $\frac{L}{\delta} - 1$, and the
  number of Steiner points corresponding to the planes $z = h(x)$ is
  at most $n-2$. So,
  \[
    |V \cap e|
    \,\le\,
    \left(\frac{L}{\delta} - 1\right) + (n-2) + 2
    \,<\,
    \frac{L}{\delta} + n
    \,=\,
    4 n \left(\frac{L}{h}\right)
         \left(\frac{1}{\epsilon}\right) \sec\theta
         \,+\, n,
  \]
  because $\delta = \frac{\epsilon h \cos\theta}{4 n}$. Since
  $\left(\frac{L}{h}\right) \left(\frac{1}{\epsilon}\right) \sec\theta
  \ge 1$, we have:
  \[
    |V \cap e|
    \,<\,
    5 n \left(\frac{L}{h}\right)
         \left(\frac{1}{\epsilon}\right) \sec\theta
    \,=\,
    \frac{5 n X}{\epsilon}
    \enspace
    .
  \]

  We will now compute $|V|$ and $|E|$. Let
  \(
    c
    =
    \frac{5 n X}{\epsilon}
  \)
  for ease of discussion. Using the fact that the number of edges is
  at most $3 n$, we have:
  \[
    |V|
    \,<\,
    3 n c
    \,=\,
    \frac{15 n^2 X}{\epsilon}
    \enspace
    .
  \]

  For each face $f$ of the terrain, there are less than $3 c$ points
  in $V \cap f$, and each such point has less than $2 c$ neighbors in
  $f$ (more precisely, in the induced subgraph $G[V \cap f]$). So, the
  number of directed links in $E$ contributed by $f$ is less than $6
  c^2$, and this bound is tight for a level face. Because there are at
  most $2 n$ faces,
  \[
    |E|
    \,<\,
    12 n c^2
    \,=\,
    O \left( \frac{n^3 X^2}{\epsilon} \right)
    \enspace
    . \qed
  \]
\end{pf}

\begin{thm} \label{thm:Steiner.ApproxAlg}
  Let $X = \left(\frac{L}{h}\right) \sec\theta$. Given a vertex $s$,
  and a constant $\epsilon \in (0,1]$, we can discretize the terrain
  with $\frac{15 n^2 X}{\epsilon}$ Steiner points so that
  after a preprocessing phase that takes
  \(
    O \left(
      \frac{n^2 X}{\epsilon} \log \left( \frac{n X}{\epsilon} \right)
    \right)
  \)
  time for a given vertex $s$, we can determine a
  $(1+\epsilon)$-approximate SDP from $s$ to any point $v$ in:
  \begin{enumerate}[(i)]
  \item $O(n)$ time if $v$ is a vertex of the terrain or a Steiner
    point, and
  \item $O \left( \frac{n X}{\epsilon} \right)$ time otherwise.
  \end{enumerate}
\end{thm}

\begin{pf}
  We first show that the path $P''$ returned by our algorithm is a
  $(1+\epsilon)$-approximation of $P$. Path $P''$ is descending
  because any path in $G$ is a descending path in the terrain, and the
  last segment of $P''$ is descending. It follows from the
  construction of $P''$ that the length of $P''$ is at most that of
  $P'$, and hence by Lemma~\ref{lem:Steiner.ApproxFactor}, $P''$ is a
  $(1+\epsilon)$-approximation of $P$.

  As we have mentioned before, we do not construct $E$ explicitly
  because the neighbors of a node $x \in V$ in the graph are
  determined during the execution of the Bushwhack algorithm. As a
  result, the (implicit) construction of $G$ takes $O(|V|)$ time. It
  follows from the running time of the Bushwhack algorithm (discussed
  in Section~\ref{L3:Back.Bushwhack}) that the preprocessing time of
  our algorithm is:
  \[
    O( |V| \log |V| )
    \,=\,
    O \left(
      \frac{n^2 X}{\epsilon} \log \left(\frac{n X}{\epsilon}\right)
    \right)
  \]
  by Lemma~\ref{lem:Steiner.GraphSize}.

  During the query phase, if $v$ is a vertex of the terrain or a
  Steiner point, the approximate path is in the tree $T$. Because the
  tree has height $O(n)$, it takes $O(n)$ time to trace the
  path. Otherwise, $v$ is an interior point of a face or an edge of
  the terrain. The last intermediate node $u$ on the path to $v$ is a
  vertex or a Steiner point that lies on the boundary of a face
  containing $v$. If $v$ is interior to a face [an edge], there are 3
  [respectively 4] edges of the terrain on which $u$ can lie. Thus
  there are $O\left(\frac{n X}{\epsilon}\right)$ choices for $u$ by
  Lemma~\ref{lem:Steiner.GraphSize}, and we try all of them to find
  the best approximate path, which takes:
  \[
    O\left(\frac{n X}{\epsilon}\right) \, + \, O(n)
    =
    O\left(\frac{n X}{\epsilon}\right)
  \]
  time. \qed
\end{pf}

Note that the space requirement of our algorithm is $O(|V|) = O \left(
\frac{n^2 X}{\epsilon} \right)$ since we are not storing $E$
explicitly. Also note that using Dijkstra's algorithm with a Fibonacci
heap~\cite{Fredman.87} instead of the Bushwhack algorithm yields an
even simpler algorithm with a preprocessing time of
\(
  O( |V| \log |V| + |E| )
  =
  O \left(
    n^3 \left(\frac{X}{\epsilon}\right)^2
  \right)
  .
\)
%

%
%
%
%

%

\section{Discretizing using Steiner Points in Geometric Progression}
\label{L2:Steiner3}

Unlike our first algorithm where the Steiner points on each edge are
evenly spaced, our second algorithm places them non-uniformly along
the edges. The Steiner points we use here are of two kinds. We first
place Steiner points in ``geometric progression'' along the edges, as
done by Aleksandrov et al.~\cite{Aleksandrov.98}. We call these points
\emph{primary Steiner points}. Then we place more Steiner points,
called \emph{isohypse Steiner points}, to guarantee that for every
descending path in the terrain there exists a descending path through
the Steiner points. Although the number of Steiner points used in this
technique is more than in our first algorithm, the running time of the
resulting algorithm no longer depends on the slope of the edges.

\subsection{Algorithm}
\label{L3:Steiner3.Alg}

\subsubsection{Preprocessing Phase}
\label{L3:Steiner3.Alg.Pre}

The primary Steiner points are placed in such a way that for each
vertex $v$ of an edge $e$, there is a set of primary Steiner points
whose distances from $v$ form a geometric progression. Although the
distance between a pair of consecutive Steiner points on $e$ increases
as we move away from $v$, we can still guarantee a good approximation
ratio. This is because intuitively the length of a segment connecting
two edges adjacent to $v$ increases as we move the segment away from
$v$---see Lemma~\ref{lem:Steiner3.FarOffset} for a more precise
statement. One observation is that if we want to maintain the
geometric progression of the distances for the Steiner points very
close to $v$, we would need infinitely many Steiner points near
$v$. To avoid this problem, we do not put any primary Steiner points
in a small region near $v$.

Before going into further details, we will define a few constants for
ease of discussion. Let
\(
  \delta_1 = \frac{\epsilon h}{6 n}
  ,
\)
and
\(
  \delta_2 = \frac{\epsilon h}{6 L}
  .
\)
The constant $\delta_1$ will define a region near $v$ where we do not
put any primary Steiner points, while $\delta_2$ will determine the
distances between consecutive primary Steiner points outside that
region.

\begin{figure}[hbt]
  \hspace*{\fill}
  \input{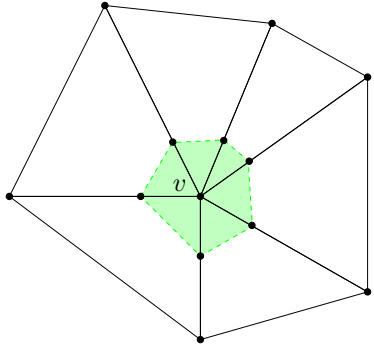}
  \hspace*{\fill}
  \caption{Vicinity of a vertex}
  \label{fig:VicinityDef}
\end{figure}

\begin{defn}[Vicinity of a Vertex]
  In a face $f$ incident to a vertex $v$, let $p_1$ and $p_2$ be two
  points lying on two different edges of $f$ at $v$ such that $|v p_1|
  = |v p_2| = \delta_1$. Clearly, $\triangle v p_1 p_2$ is an
  isosceles triangle. The \emph{vicinity of $v$\/} is defined to be
  the union of all such isosceles triangles around $v$
  (Figure~\ref{fig:VicinityDef}).
\end{defn}

Note that the vicinities of any two vertices $v_1$ and $v_2$ are
mutually disjoint because $\delta_1 < \frac{h}{2} < \frac{|v_1
v_2|}{2}$.

In the preprocessing phase, we determine the positions of the Steiner
points as follows. First, on every edge $e = v_1 v_2$ we place primary
Steiner points at points $p \in e$ such that $|p q| = \delta_1 (1 +
\delta_2)^i$ for $q \in \{v_1,v_2\}$ and $i \in
\{0,1,2,\ldots\}$. Then we add up to $3 n$ isohypse Steiner points for
each primary Steiner point and for each vertex, as follows. For every
non-level edge $e$, and every point $p$ that is either a primary
Steiner point or a vertex, we place an isohypse Steiner point at the
point where $e$ intersects the horizontal plane through $p$ (i.e.,~the
plane $z = h(p)$).

After placing the Steiner points, we construct a weighted directed
graph $G=(V,E)$ and then construct a shortest path tree $T$ rooted at
$s$ in $G$ in the same way as in our first algorithm
(Section~\ref{L3:Steiner.Alg.Pre}).

\subsubsection{Query Phase}
\label{L3:Steiner3.Alg.Qry}

The queries are handled in exactly the same manner as in
Section~\ref{L3:Steiner.Alg.Qry}.

\subsection{Correctness and Analysis}
\label{L3:Steiner3.Analysis}

For the proof of correctness, we follow the same approach used in
Section~\ref{L3:Steiner.Analysis}: given an SDP $P$, we first
construct a path $P'$ by moving each intermediate node of $P$ upward
to the nearest Steiner point, and then show that $P'$ is descending
and that it approximates $P$. This proves the correctness of our
algorithm because the path returned by our algorithm is not longer than
$P'$.

Let $P = (s=p_0, p_1, p_2, \ldots, p_k, v=p_{k+1})$ be an SDP from $s$
to $v$ such that $p_i$ and $p_{i+1}$ are two different boundary points
of a common face for all $i \in [0,k-1]$, and $p_k$ and $p_{k+1}$ are
two points of a common face. Let $e_i$ be an edge of the terrain
through $p_i$ for all $i \in [1,k]$; $e_i$ can be any edge through
$p_i$ if $p_i$ is a vertex. Now define path $P' = (s=p'_0, p'_1, p'_2,
\ldots, p'_k, v=p'_{k+1})$ as follows: for each $i \in [1,k]$, let
$p'_i = p_i$ if $p_i$ is a vertex of the terrain; otherwise, let
$p'_i$ be the nearest point from $p_i$ in $V \cap e_i$ such that
$h(p'_i) \ge h(p_i)$.

\begin{lem} \label{lem:Steiner3.Feasibility}
  Path $P'$ is descending, and the part of $P'$ from $s$ to $p'_k$
  exists in $G$.
\end{lem}

\begin{pf}
  The proof is exactly the same as in
  Lemma~\ref{lem:Steiner.Feasibility}. \qed
\end{pf}

\begin{lem} \label{lem:Steiner3.FarOffset}
  For all $i \in [1,k]$ such that $p_i$ is not inside a vertex
  vicinity,
  \[
    |p_i p'_i| < \frac{\epsilon}{6} \, |p_{i-1} p_i|
    \enspace .
  \]
\end{lem}

\begin{figure}[bht]
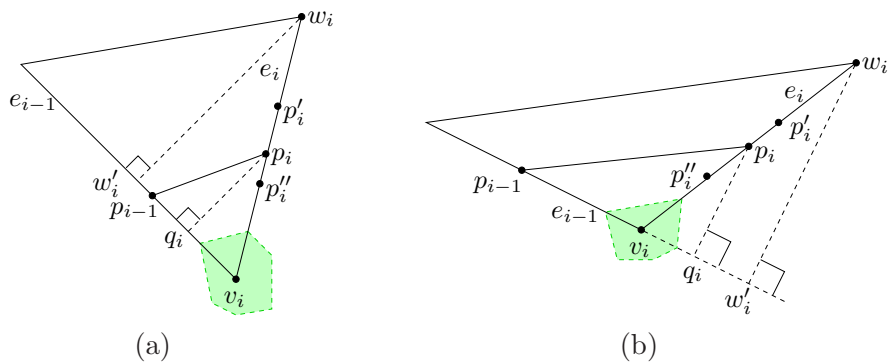

  \hspace*{\fill}
  \subfigure[]{
    \input{fig/GeometricStPts1.pstex_t}
    \label{fig:GeometricStPts1}
  }
  \hspace*{\fill}
  \subfigure[]{
    \input{fig/GeometricStPts2.pstex_t}
    \label{fig:GeometricStPts2}
  }
  \hspace*{\fill}
  \caption{Bounding $|p_i p'_i|$ when the face angle at $v_i$ is (a)
    acute and (b) obtuse}
  \label{fig:GeometricStPts}
\end{figure}

\begin{pf}
  If $p_i$ coincides with $p'_i$, the lemma follows trivially as $|p_i
  p'_i| = 0$. We will now focus on the case when these two points do
  not coincide. Since $p_i$ is not inside a vertex vicinity, there is
  another Steiner point $p''_i$ in $e_i$ such that $p'_i$ and $p''_i$
  lie on the opposite sides of $p_i$. Let $v_i$ be the common vertex
  of $e_{i-1}$ and $e_i$, $w_i$ be the other vertex of $e_i$, and
  $q_i$ and $w'_i$ be two points in $e_{i-1}$ such that $p_i
  q_i \perp e_{i-1}$ and $w_i w'_i \perp
  e_{i-1}$. Figure~\ref{fig:GeometricStPts} depicts these vertices and
  points, for both the cases that the face angle at $v_i$ is (a)~acute
  and (b)~obtuse.

  We will first show that $|p'_i p''_i| < \delta_2 |v_i p_i|$, and
  then prove the lemma using a property of similar triangles. We have
  two cases as follows. If $|v_i p''_i| < |v_i p'_i|$, then by
  construction:
  \begin{eqnarray*}
    &&
    |v_i p'_i|
    \;\le\;
    (1 + \delta_2) \, |v_i p''_i|
    \\
    &\Rightarrow&
    |v_i p'_i| - |v_i p''_i|
    \;\le\;
    \delta_2 \, |v_i p''_i|
    \\
    &\Rightarrow&
    |p'_i p''_i|
    \;\le\;
    \delta_2 \, |v_i p''_i|
    <
    \delta_2 \, |v_i p_i|
    \enspace ,
  \end{eqnarray*}
  since $p_i$ lies strictly in between $p'_i$ and $p''_i$. On the
  other hand, if $|v_i p''_i| > |v_i p'_i|$, then by construction:
  \begin{eqnarray*}
    &&
    |v_i p''_i|
    \;\le\;
    (1 + \delta_2) \, |v_i p'_i|
    \\
    &\Rightarrow&
    |v_i p''_i| - |v_i p'_i|
    \;\le\;
    \delta_2 \, |v_i p'_i|
    \\
    &\Rightarrow&
    |p'_i p''_i|
    \;\le\;
    \delta_2 \, |v_i p'_i|
    \;<\;
    \delta_2 \, |v_i p_i|
    \enspace ,
  \end{eqnarray*}
  since $p_i$ lies strictly in between $p'_i$ and $p''_i$. In both
  cases, $|p'_i p''_i| < \delta_2 |v_i p_i|$.

  We have:
  \begin{eqnarray*}
    |p_i p'_i|
    &<&
    |p'_i p''_i|
    \;<\;
    \delta_2 \, |v_i p_i|
    \\
    &=&
    \delta_2 \, |q_i p_i| \cdot \frac{|v_i p_i|}{|q_i p_i|}
    \\
    &=&
    \delta_2 \, |q_i p_i| \cdot \frac{|v_i w_i|}{|w'_i w_i|}
    \qquad
    \mbox{(since $\triangle v_i p_i q_i$ and $\triangle v_i w_i
      w'_i$ are similar)}
    \\
    &\le&
    \delta_2 \, |q_i p_i| \cdot \frac{L}{h}
    \\
    &=&
    \frac{\epsilon h}{6 L} \cdot |q_i p_i| \cdot \frac{L}{h}
    \qquad
    \mbox{(from the definition of $\delta_2$)}
    \\
    &\le&
    \frac{\epsilon}{6} \, |p_{i-1} p_i|
    \enspace
    \qquad
    \mbox{(since $|p_{i-1} p_i| \ge |q_i p_i|$)}
    \enspace
    . \qed
  \end{eqnarray*}
\end{pf}

\begin{lem} \label{lem:Steiner3.NearOffset}
  For all $i \in [1,k]$ such that $p_i$ is on or inside a vertex
  vicinity,
  \[
    |p_i p'_i| \le \frac{\epsilon h}{6 n}
    \enspace .
  \]
\end{lem}

\begin{pf}
  If $p_i$ is a vertex, the lemma follows trivially since $p'_i = p_i$
  in this case. If $p_i$ is not a vertex, let $e_i$ be the edge
  containing $p_i$, and $v_i$ be the vertex whose vicinity contains
  $p_i$. It is not hard to see that $v_i$ is a vertex of $e_i$ because
  $\delta_1$ is strictly less than $h$. Let $q_i$ be the primary
  Steiner point on $e_i$ which lies at distance $\delta_1$ from
  $v_i$. Clearly $p_i$ lies in line segment $v_i q_i$. Now $p'_i$
  cannot be outside line segment $v_i q_i$ because otherwise we would
  have chosen either $v_i$ or $q_i$ as $p'_i$. As a result, $p'_i$
  also lies in line segment $v_i q_i$. Therefore,
  \[
    |p_i p'_i| \le |v_i q_i| = \delta_1 = \frac{\epsilon h}{6 n}
    \enspace
    . \qed
  \]
\end{pf}

\begin{lem} \label{lem:Steiner3.ApproxFactor}
  Path $P'$ is a $(1+\epsilon)$-approximation of $P$.
\end{lem}

\begin{pf}
  The length of $P'$ is equal to:
  \begin{eqnarray*}
    \sum_{i=0}^k |p'_i p'_{i+1}|
    &\le&
    \sum_{i=0}^k \left(
      |p'_i p_i| + |p_i p_{i+1}| + |p_{i+1} p'_{i+1}|
    \right)
    \qquad
    \mbox{(from triangle inequality)}
    \\
    &=&
    \sum_{i=0}^k |p_i p_{i+1}| \; + \; 2 \sum_{i=1}^k |p_i p'_i|
    \qquad
    \mbox{(since $p_0 = p'_0$ and $p_{k+1} = p'_{k+1}$)}
    \\
    &<&
    \sum_{i=0}^k |p_i p_{i+1}| \; + \;
      2 \sum_{i=1}^k \left(
        \frac{\epsilon}{6} \, |p_{i-1} p_i| + \frac{\epsilon h}{6 n}
      \right)
    \qquad
    \mbox{(by Lemmas~\ref{lem:Steiner3.FarOffset}
      and~\ref{lem:Steiner3.NearOffset})}
    \\
    &=&
    \sum_{i=0}^k |p_i p_{i+1}| \; + \;
      \frac{\epsilon}{3} \sum_{i=1}^k |p_{i-1} p_i| \; + \;
      \frac{\epsilon h k}{3 n}
    \\
    &\le&
    \sum_{i=0}^k |p_i p_{i+1}| \left( 1 + \frac{\epsilon}{3} \right) \; + \;
      \frac{\epsilon h k}{3 n}
    \\
    &<&
    \sum_{i=0}^k |p_i p_{i+1}| \left( 1 + \frac{\epsilon}{3} \right) \; + \;
      \frac{2 \epsilon h}{3}
    \enspace ,
  \end{eqnarray*}
  since $k < 2 n$ because the number of faces in the terrain is at
  most $2 n$, and $P$ has at most one segment in each face
  (Lemma~\ref{lem:SDP.OneLinePerFace}). Assuming that $P$ crosses at
  least one edge of the terrain (otherwise, $P' = (s,v) = P$),
  $\sum_{i=0}^k |p_i p_{i+1}| \ge h$, and therefore:
  \begin{eqnarray*}
    \sum_{i=0}^k |p'_i p'_{i+1}|
    &<&
    \sum_{i=0}^k |p_i p_{i+1}| \left( 1 + \frac{\epsilon}{3} \right) \; + \;
      \frac{2 \epsilon}{3} \sum_{i=0}^k |p_i p_{i+1}|
    \\
    &=&
    \left(
      1 + \frac{\epsilon}{3} +
        \frac{2 \epsilon}{3}
    \right)
    \sum_{i=0}^k |p_i p_{i+1}|
    \\
    &=&
    (1 + \epsilon) \, \sum_{i=0}^k |p_i p_{i+1}|
    \enspace .
  \end{eqnarray*}
  Because $P'$ is descending (Lemma~\ref{lem:Steiner3.Feasibility}),
  it follows that $P'$ is a $(1+\epsilon)$-approximation of $P$. \qed
\end{pf}

\begin{obs} \label{obs:Steiner3.LogMath}
  For any real number $x \in (0,1]$, $\log(1+x) > \frac{x \log e}{2}$.
\end{obs}

\begin{pf}
  \begin{eqnarray*}
    \log(1+x)
    &=&
    \log e \cdot \log_e (1+x)
    \\
    &=&
    \log e \left(
      x - \frac{x^2}{2} +
      \frac{x^3}{3} - \frac{x^4}{4} +
      \frac{x^5}{5} - \frac{x^6}{6} +
      \ldots
    \right)
    \\
    &=&
    \log e \left(
      x   \left(1 - \frac{x}{2}\right) +
      x^3 \left(\frac{1}{3} - \frac{x}{4}\right) +
      x^5 \left(\frac{1}{5} - \frac{x}{6}\right) +
      \ldots
    \right)
    \enspace .
  \end{eqnarray*}
  Since each term in the outer parentheses of the last expression is
  strictly positive, we have:
  \begin{eqnarray*}
    \log(1+x)
    &>&
    \log e \left(
      x   \left(1 - \frac{x}{2}\right)
    \right)
    \;\ge\;
    \log e \left(
      x   \left(1 - \frac{1}{2}\right)
    \right)
    \;=\;
    \frac{x \log e}{2}
    \enspace
    . \qed
  \end{eqnarray*}
\end{pf}

\begin{lem} \label{lem:Steiner3.GraphSize}
  Graph $G$ has less than
  \(
    \frac{153 n^2 L}{\epsilon h}
      \log \left( \frac{6 n L}{\epsilon h} \right)
  \)
  nodes and
  \(
    O \left( \frac{n^3 L^2}{\epsilon^2 h^2}
      \log^2 \left( \frac{n L}{\epsilon h} \right)
    \right)
  \)
  links. Moreover, it has less than
  \(
    \frac{51 n L}{\epsilon h}
      \log \left( \frac{6 n L}{\epsilon h} \right)
  \)
  nodes along any edge of the terrain.
\end{lem}

\begin{pf}
  We will first compute an upper bound on the number of primary
  Steiner points, which will then be used to prove the lemma.

  Let $n_e$ be the number of primary Steiner points on edge $e$. It is
  straightforward to see that $n_e$ is at most $2 j$, where $j$ is the
  largest integer satisfying the following inequality:
  \begin{eqnarray*}
    &&
    \delta_1 (1 + \delta_2)^j
    <
    L
    \\
    &\Rightarrow&
    (1 + \delta_2)^j
    <
    \frac{L}{\delta_1}
    \\
    \\
    &\Rightarrow&
    j
    <
    \frac{\log \left( \frac{L}{\delta_1} \right)}{\log(1 + \delta_2)}
    \enspace
    .
  \end{eqnarray*}
  Therefore,
  \begin{eqnarray*}
    n_e
    &\le&
    2 j
    \;<\;
    \frac{2 \log \left( \frac{L}{\delta_1} \right)}
             {\log(1 + \delta_2)}
    \\
    &<&
    \frac{2}{\frac{\delta_2 \log e}{2}}
      \, \log \left( \frac{L}{\delta_1} \right)
    \qquad
    \mbox{(Lemma~\ref{obs:Steiner3.LogMath})}
    \\
    &=&
    \frac{4 \log_e 2 \cdot 6 L}{\epsilon h}
      \log \left( \frac{L \cdot 6 n}{\epsilon h} \right)
    \\
    &<&
    \frac{16.64 L}{\epsilon h}
      \log \left( \frac{6 n L}{\epsilon h} \right)
    \enspace
    .
  \end{eqnarray*}
  Since there are at most $3 n$ edges in the terrain, the total number
  of primary Steiner points is at most $3 n n_e$, which is less than:
  \[
    \frac{50 n L}{\epsilon h}
      \log \left( \frac{6 n L}{\epsilon h} \right)
    \enspace .
  \]

  We will now prove the last part of the lemma. For each point $p$
  that is either a primary Steiner points and a vertex, there is at
  most one node in $V \cap e$ for any edge $e$. This is obvious when
  $p$ lies on $e$. On the other hand, if $p$ does not lie on $e$,
  there is at most one isohypse Steiner point on $e$ that corresponds
  to $p$. Using the above bound on the number of primary Steiner
  points, we have:
  \begin{eqnarray*}
    |V \cap e|
    &<&
    \frac{50 n L}{\epsilon h}
      \log \left( \frac{6 n L}{\epsilon h} \right)
    \, + \,
    n
    \\
    &<&
    \frac{51 n L}{\epsilon h}
      \log \left( \frac{6 n L}{\epsilon h} \right)
    \enspace
    \qquad
    \mbox{(since $\frac{L}{h} > 1$).}
  \end{eqnarray*}

  We will now compute $|V|$ and $|E|$. Let
  \(
    c
    =
    \frac{51 n L}{\epsilon h}
      \log \left( \frac{6 n L}{\epsilon h} \right)
  \)
  for ease of discussion. Using the fact that the number of edges is
  at most $3 n$, we have:
  \[
    |V|
    \,<\,
    3 n c
    \,=\,
    \frac{153 n^2 L}{\epsilon h}
      \log \left( \frac{6 n L}{\epsilon h} \right)
    \enspace
    .
  \]

  Using the same argument we used in the proof of
  Lemma~\ref{lem:Steiner.GraphSize}, we can say that the number of
  directed links in $E$ contributed by each $f$ of the terrain is less
  than $6 c^2$. Because there are at most $2 n$ faces,
  \[
    |E|
    \,<\,
    12 n c^2
    \,=\,
    O \left( \frac{n^3 L^2}{\epsilon^2 h^2}
      \log^2 \left( \frac{n L}{\epsilon h} \right)
    \right)
    \enspace
    . \qed
  \]
\end{pf}

\begin{thm} \label{thm:Steiner3.ApproxAlg}
  Given a vertex $s$, and a constant $\epsilon \in (0,1]$, we can
  discretize the terrain with at most
  \(
    \frac{150 n^2 L}{\epsilon h}
      \log \left( \frac{6 n L}{\epsilon h} \right)
  \)
  Steiner points so that after a preprocessing phase that takes
  \(
    O \left(
      \frac{n^2 L}{\epsilon h}
      \log^2 \left( \frac{n L}{\epsilon h} \right)    
    \right)
  \)
  time for a given vertex $s$, we can determine a
  $(1+\epsilon)$-approximate SDP from $s$ to any point $v$ in:
  \begin{enumerate}[(i)]
  \item $O(n)$ time if $v$ is a vertex of the terrain or a Steiner
    point, and
    \item 
    \(
      O \left(
        \frac{n L}{\epsilon h}
          \log \left( \frac{n L}{\epsilon h} \right)
      \right)
    \)
    time otherwise.
  \end{enumerate}
\end{thm}

\begin{pf}
  The proof is the same as in Theorem~\ref{thm:Steiner3.ApproxAlg}
  except that we use Lemmas~\ref{lem:Steiner3.ApproxFactor}
  and~\ref{lem:Steiner3.GraphSize} instead of
  Lemmas~\ref{lem:Steiner.ApproxFactor}
  and~\ref{lem:Steiner.GraphSize} respectively.
\end{pf}

As in the case of our first algorithm, we can use Dijkstra's algorithm
with a Fibonacci heap~\cite{Fredman.87} instead of the Bushwhack
algorithm to have an even simpler algorithm with a preprocessing time
of
\(
  O \left( \frac{n^3 L^2}{\epsilon^2 h^2}
      \log^2 \left( \frac{n L}{\epsilon h} \right)
    \right)
  .
\)
%

%
%
%
%

%

\section{Conclusion}
\label{L2:Conclusion}

It may appear that the running time can be improved by using the
technique by Aleksandrov et al.~\cite{Aleksandrov.05} who place
Steiner points along the bisectors of the face angles. Although the
technique improves all previous results on the Weighted Region
Problem, it cannot be used for the SDP problem very easily. The main
problem is that it is not clear how to prove the existence of a
\emph{feasible\/} path that approximates an SDP.

When query point $v$ is neither a vertex of the terrain nor a Steiner
point, the query phase can be made faster by using a point location
data structure on each face. Note that the Voronoi diagram on each
face consists of hyperbolic arcs.

\bibliographystyle{plain}
\bibliography{Refs}

\end{document}